\documentclass[epj]{svjour}

\usepackage{graphicx}
\usepackage{fancyhdr}
\def\makeheadbox{{
 }}
\def\mytitle{My title} 
\def\myauthors{My name}  
\def\mytype{My type of session}
\def\mysession{My session}

\def\mytitle{Flavour symmetries and SUSY soft breaking in the LHC era} 
\def\myauthors{Oscar Vives}    
\def\mytype{Contributed Talk}    
\def\mysession{Flavor Physics}

\newcommand{\Frac}[2]{\frac{\displaystyle{#1}}{\displaystyle{#2}}}

\newcommand{\bea}{\begin{eqnarray}}
\newcommand{\beq}{\begin{equation}}
\newcommand{\eea}{\end{eqnarray}}
\newcommand{\eeq}{\end{equation}}
\newcommand{\unity}{{\hbox{1\kern-.8mm l}}}

\begin{document}
\title{Flavour symmetries and SUSY soft breaking in the LHC era}
\author{Oscar Vives\inst{1}
\thanks{\emph{Email:} oscar.vives@uv.es}%
}                     
%
%
\institute{Departament de F\'{\i}sica Te\`orica and IFIC, Universitat de 
Valencia-CSIC, 
46100 Valencia, Spain
}
%
\date{}
\abstract{
The so-called  supersymmetric flavour problem does not exist in isolation
to the  Standard Model flavour problem. We show that a realistic flavour
symmetry can simultaneously solve both problems without ad hoc modifications
of the SUSY model. Furthermore, departures from the SM expectations in these
models  can be used to discriminate among different possibilities. 
In particular we present the
expected values for the electron EDM in a flavour model solving
the supersymmetric flavour and CP problems.
\PACS{{12.60.Jv}{Supersymmetric models} \and {12.15.Ff}{Quark and lepton masses and mixing}
     } 
} 
\maketitle
%
\section{Introduction}

The SUSY flavour problem has been traditionally used to justify different
departures from the ``natural'' gravity-mediated MSSM setting. However, 
in this talk we will take a different point of view
and we will show that the so-called {\it supersymmetric flavour problem} 
does not really exist, or more exactly, it can not be detached from the 
{\it Standard Model flavour problem}. In fact a correct solution to the
Standard Model flavour problem will probably pass unscathed all the stringent
constraints on flavour changing neutral currents after the inclusion of the
MSSM soft sector.  
   
The supersymmetric flavour problem is usually stated as follows: The SUSY 
soft-breaking terms and they have a completely different origin from the 
Yukawa couplings in the superpotential and we have no information on their
structure.  In principle,
we could expect that all the entries in the soft breaking matrices were $O(1)$
in any basis and in particular in the basis where the Yukawa couplings are
diagonal. In this situation FCNC and $CP$ violation observables would receive
too large contributions from loops involving SUSY particles and this disagrees
strongly with the stringent phenomenological bounds on these processes.
As formulated above we can only agree with this statement, however, it is
trivial to reformulate this statement in terms of the Yukawa couplings of the
superpotential: We have no theoretical guidance to build the Yukawa
couplings. If we had to write an SM Lagrangian ignoring the measured quark 
and lepton masses and mixings, any
flavour structure would be possible and in fact we would naturally expect all
the different entries in the Yukawa matrices to be $O(1)$. Clearly this would
never agree with the observed fermion masses and mixing angles.   
Therefore we have to conclude that there is a much stronger flavour problem in
the SM than in the MSSM. The real {\bf flavour problem} is simply our
inability to understand the complicated structures in the quark and lepton
Yukawa couplings and likewise soft-breaking flavour structures in the MSSM.

At this point we have to emphasize that the presence of new physics, as 
for instance supersymmetry, is not a problem for flavour but on the contrary
a necessary tool to advance in our understanding of the flavour problem. 
In the framework of the Standard Model all the information we can extract on
flavour are the Yukawa eigenvalues (quark and lepton masses) and the 
left-handed misalignment between up and down quarks (CKM matrix) or leptons 
(MNS matrix) and this is not enough to determine the full structure of the
Yukawa matrices. However, in supersymmetric extensions of the SM, the new 
interactions can 
provide additional information on the physics of flavour which will be 
fundamental to improve our knowledge on flavour.
In the following we show that finding a solution to the ``SM'' flavour
problem will also solve the so-called ``supersymmetric
flavour problem'' to a sufficient degree.

\section{Flavour symmetries}
The flavour structure associated to the SM Yukawas is 
very special: a strong hierarchy in the couplings and a peculiar structure 
of the mixing matrices. In a truly fundamental theory we would expect all 
dimensionless couplings to be $O(1)$ and thus these small couplings must be
explained. The basic idea of flavour symmetries is to use an spontaneously 
broken family symmetry in analogy with the gauge sector to
generate these couplings. A scalar vev
breaking the flavour symmetry normalized with a large mediator mass provides
a small expansion parameter that enters in different powers in the fermion 
Yukawa couplings\cite{Froggatt:1978nt}. In the limit of exact
symmetry the Yukawa couplings are forbidden and only when the symmetry is
broken these couplings appear as a function of small vevs. Similarly,  
in a supersymmetric theory, the flavour symmetry applies both to the fermion 
and sfermion sectors. Therefore, the structures in the 
soft-breaking matrices and the Yukawa couplings are related.  
The starting point in our analysis is then the texture in the Yukawa 
couplings. However,  the complete texture of the Yukawa matrices can not be 
fixed though Standard Model interactions. 
Still, it is reasonable to assume that the smallness of CKM mixing angles 
is due to the smallness of the off-diagonal elements in the Yukawa matrices 
with respect to the corresponding diagonal elements. Then we can fix
the elements above the diagonal, corresponding to the left-handed mixings, but
not the elements below the diagonal \cite{Roberts:2001zy}. 
Therefore, we can consider two complementary situations that we call symmetric
and asymmetric Yukawa textures. In the symmetric textures we make the 
additional simplifying assumption of choosing the matrices to be
symmetric. Note that this situation is not unusual in many flavour models
\cite{King:2001uz,King:2003rf,Ross:2004qn} as well as in GUT theories.
Asymmetric textures are also common in simple Abelian 
flavour symmetries with a single flavon field \cite{Leurer:1992wg,Dudas:1995yu}.

The simplest example is provided by a $U(1)$ flavour symmetry, as originally
considered by Froggatt and Nielsen \cite{Froggatt:1978nt}, which generates
an asymmetric texture. As an example we can assign the three generations of SM 
fields the charges: $Q_i=(3,2,0)$, $d_i^c=(0,0,1)$, $u_i^c=(3,2,0)$ with
a single flavon field of charge $-1$. The vev of the
flavon field normalized to the mass of the heavy mediator fields
$M_f$, is  $\epsilon = v/M_f \ll 1$. The superpotential of this model is:
\begin{eqnarray} 
W_{\rm Y} = Q_i d^c_j H_1 \left(\frac{\theta}{M_{\rm fl}}\right)^{q_i+d_j}
+ Q_i u^c_j H_2 \left(\frac{\theta}{M_{\rm fl}}\right)^{q_i+u_j}
\end{eqnarray} 
where unknown $O(1)$ coefficients have been suppressed for clarity. 
Then we have,
\begin{equation}
Y_u= \left( \matrix{ \epsilon^6&\epsilon^5&\epsilon^3 \cr \epsilon^5&\epsilon^4&\epsilon^2 \cr \epsilon^3&\epsilon^2&1 }  \right) 
~~~~,~~~~
Y_d = \left( \matrix{ \epsilon^4&\epsilon^3&\epsilon^3 \cr \epsilon^3&\epsilon^2&\epsilon^2 \cr \epsilon&1&1 }  \right) 
~~~~,~~~~
\end{equation}
The soft masses are couplings $\phi^{\dagger }\phi$, clearly invariant under
any symmetry,  and therefore always allowed. Hence, diagonal soft masses are 
allowed in the limit of unbroken symmetry and unsuppressed. 
Assuming diagonal masses of different generations are equal in the symmetric
limit\footnote{Unlike
  in the case of non-Abelian symmetries, this is not guaranteed by symmetry, 
  but it is still possible  in some cases like dilaton domination in gravity
  mediation models.}, the universality is then broken by the flavon vevs. 
Any combination of two MSSM scalar fields $\phi_i$ and an arbitrary number of 
flavon vevs invariant under the symmetry will contribute to the soft masses:
\begin{eqnarray}
{\cal L}_{m^2}& =& m_0^2 \left(\phi_1^* \phi_1 + \phi_2^* \phi_2 + \phi_3^*
  \phi_3 \right. \nonumber \\ &+& \left.\left(\frac{\langle\theta\rangle}{M_{\rm
  fl}}\right)^{q_j-q_i}~ \phi_i^* \phi_j  + {\rm h.c.} \right).
\end{eqnarray}
Thus, the structure of the right-handed down squark mass matrix we would 
have in this model is:    
\begin{eqnarray}
M^2_{\tilde{D}_R} \simeq \left(
\begin{array}{ccc} 1 & {\epsilon} & {\epsilon}  \\ 
{\epsilon} & 
1 & 1\\
{\epsilon} & 1 & 1  \end{array}
\right) m_0^2 \, .
\end{eqnarray}
In this case, we would expect large mixings in the second and third
generations of right-handed down sfermions. Notice however, that this simple
model is already ruled-out by the stringent constraints in the 1--2 sector
unless sfermions are very heavy.

Symmetric textures are obtained, for instance, from a spontaneously broken 
$SU(3)$ family symmetry. The basic features of this symmetry are the
following. All left handed fermions ($\psi_i$ and $\psi^c_i$) are triplets
under $SU(3)_{fl}$. To allow for the spontaneous symmetry breaking of 
$SU(3)$ it is necessary to
add several new scalar fields which are either triplets ($\overline{\theta}%
_{3}$, $\overline{\theta}_{23}$, $\overline{\theta}_{2}$) or antitriplets ($%
\theta_{3}$, $\theta_{23}$). We assume that $SU(3)_{fl}$ is broken in two
steps. The first step occurs when $\theta_3$ and $\bar \theta_{3}$ get
a large vev breaking $SU(3)$ to $SU(2)$. Subsequently a smaller vev of 
$\theta_{23}$ and $\bar \theta_{23}$ breaks the
remaining symmetry. After this breaking we obtain the effective Yukawa
couplings through the Froggatt-Nielsen mechanism \cite{Froggatt:1978nt}
integrating out heavy fields. In fact, to reproduce measured masses and
mixings, the large third generation Yukawa
couplings require  a $\theta_3, \bar \theta_{3}$ vev of the order of the
mediator scale, $M_f$, 
while $\theta_{23}/M_f, \bar \theta_{23}/M_f$ have
vevs of order $\varepsilon=0.05$ in the up sector and $\bar \varepsilon=0.15$ 
in the down sector with different mediator scales in both sectors.
Moreover in the minimization of the scalar potential it is possible to
ensure that the fields $\theta_{23}$ and $\bar \theta_{23}$ get equal 
vevs in the second and third components. In this model, CP is spontaneously
broken by the flavon vevs that are complex generating the observed CP
violation in the CKM matrix.
The basic structure of the Yukawa superpotential is then given by:
\begin{eqnarray}
W_{\rm Y} &=& H\psi _{i}\psi _{j}^{c} \left[ \theta _{3}^{i}\theta
_{3}^{j}+\theta _{23}^{i}\theta _{23}^{j}\right.\nonumber\\&+&\left.
\epsilon ^{ikl}%
\overline{\theta }_{23,k}\overline{\theta }_{3,l}\theta _{23}^{j}\left(
\theta _{23}\overline{\theta _{3}}\right) +\dots \right].
\end{eqnarray} 
This structure is quite general for the different $SU(3)$ models we can
build, for additional details we refer to 
\cite{Ross:2004qn,King:2001uz,King:2003rf}. The Yukawa textures are 
then symmetric and suppressing $O(1)$ coefficients:  
\begin{eqnarray}  
\label{fit}
Y_d\propto\left( 
\begin{array}{ccc}
0 & \bar \varepsilon^{3} & {\ \bar \varepsilon^{3}} \\ 
\bar \varepsilon^{3} & {\bar \varepsilon^{2}} & 
{\ \bar \varepsilon^{2}} \\ 
{\ \bar \varepsilon^{3}} & {\ \bar \varepsilon^{2}} & 1%
\end{array}%
\right),~~~~~~ Y_u\propto \left( 
\begin{array}{ccc}
0 & {\ \varepsilon^{3}} & {\ \varepsilon^{3}} \\ 
{\ \varepsilon^{3}} & {\ \varepsilon^{2}} & 
\varepsilon^{2} \\ 
{\ \varepsilon^{3}} & \varepsilon^{2} & 1%
\end{array}
\right ) \, .
\end{eqnarray}

In the same way after  $SU(3)$ breaking the scalar soft masses deviate 
from  exact universality. In first place we must notice that a mass 
term $\psi _{i}^{\dagger }\psi _{i}$ is invariant under any symmetry and 
hence gives rise to a
common contribution for the family triplet. However, $SU(3)$ breaking 
terms give rise to
important corrections \cite{Ross:2004qn,Ross:2002mr}. Any invariant
combination of flavon fields can also contribute to the sfermion
masses. Including these corrections the leading contributions 
to the sfermion mass matrices are:
\begin{eqnarray}
(M^2_{\tilde f})^{ij}&=& m_0^2\left(\delta ^{ij} +\frac{\displaystyle{1}}{\displaystyle{M_f^{2}}}\left[\theta _{3}^{i\dagger
}\theta _{3}^{j} +\theta _{23}^{i\dagger }\theta_{23}^{j}\right]\right. \nonumber \\
&+&
\left.\Frac{1}{M_f^4}(\epsilon ^{ikl}\overline{\theta }_{3,k}
\overline{\theta }_{23,l})^{\dagger }(\epsilon ^{jmn}
\overline{\theta }_{3,m}\overline{\theta }_{23,n})\right) ,
\end{eqnarray}
where $f$ represents the $SU(2)$ doublet or the up and down singlets with 
$M_f=M_L, M_u, M_d$. For instance, the down squark and charged slepton mass 
matrices after 
running to the electroweak scale and in the basis of diagonal charged lepton
Yukawas (the so-called SCKM basis) are, 
\begin{eqnarray}
&{M^2_{\tilde{D}_R}} \simeq  
6~{ M_{1/2}^2}~{\bf \unity}\qquad\qquad\qquad\qquad\qquad\qquad \\&+\left(
\begin{array}{ccc} 1 + { \bar \varepsilon^3} &  
{\bar\varepsilon^3} & {\bar\varepsilon^3} \\ 
{\bar\varepsilon^3} & 
1 + {\bar\varepsilon^2} & {\bar \varepsilon^2} \\
{\bar\varepsilon^3} & {\bar\varepsilon^2} & 1 + {\bar
\varepsilon} \end{array}
\right) { m_0^2} \nonumber \\
&{M^2_{\tilde{D}_L}}~\simeq~6~{ M_{1/2}^2}~{\bf \unity}\qquad\qquad\qquad\qquad\qquad\qquad\\
&+\left(
\begin{array}{ccc} 1 + { \varepsilon^3} &  
{\varepsilon^2 \bar \varepsilon} & {\varepsilon^2 \bar \varepsilon}+ {c_{\rm run}}~{\bar \varepsilon^3} \\ 
{\varepsilon^2 \bar \varepsilon} & 
1 + {\varepsilon^2} & {\varepsilon^2} + {c_{\rm run}}~{\bar \varepsilon^2} \\
{\varepsilon^2 \bar \varepsilon}+ {c_{\rm run}}~{\bar
  \varepsilon^3}  & {\varepsilon^2} + {c_{\rm run}}~{\bar \varepsilon^2} & 1 + {\bar
\varepsilon} \end{array}
\right) { m_0^2} \nonumber \\
&{M^2_{\tilde{E}_R}} \simeq  
0.15~{M_{1/2}^2}~{\bf \unity}\qquad\qquad\qquad\qquad\qquad\qquad
\\&~~~~~ + \left(
\begin{array}{ccc} 1 + {\bar \varepsilon^3} &\frac{\bar\varepsilon^3}{3}e^{i \alpha} & {\bar\varepsilon^3} e^{i \beta}\\ 
\frac{\bar\varepsilon^3}{3} e^{-i \alpha}& 
1 + {\bar\varepsilon^2} & {\bar \varepsilon^2} e^{i \omega} \\
{\bar\varepsilon^3} e^{-i \beta}& {\bar\varepsilon^2}e^{-i \omega} & 1 + {\bar
\varepsilon} \end{array}
\right) {m_0^2} \nonumber \\
&{M^2_{\tilde{E}_L}}~\simeq 0.5~{ M_{1/2}^2}~{\bf
  \unity}\qquad\qquad\qquad\qquad\qquad\qquad
\\&~~~~+\left(
\begin{array}{ccc} 1 + { \varepsilon^3} &  
\frac{\varepsilon^2 \bar \varepsilon}{3} & {\varepsilon^2 \bar \varepsilon}+ {c_{\rm run}}~{\bar \varepsilon^3} \\ 
\frac{\varepsilon^2 \bar \varepsilon}{3} & 
1 + {\varepsilon^2} & {\varepsilon^2} + 3 {c_{\rm run}}~{\bar \varepsilon^2} \\
{\varepsilon^2 \bar \varepsilon}+ {c_{\rm run}}~{\bar
  \varepsilon^3}  & {\varepsilon^2} + 3 {c_{\rm run}}~{\bar \varepsilon^2} & 1 + {\bar
\varepsilon} \end{array}
\right) { m_0^2} \, ,\nonumber
\label{soft2}
\end{eqnarray} 
where we include a contribution from the RGE evolution of the sfermion masses
with a coefficient $c_{\rm run}$ typically of order $0.1$, which in these
cases is more
important than the ``tree level'' contributions. Therefore we can see that the
``natural'' structures in the soft mass matrices for the 
symmetric Yukawas are different from tose in the asymmetric 
case and this provides a chance to distinguish the two Yukawa structures
through an analysis of the flavour structures in the soft SUSY sector. 

\begin{figure}
\includegraphics[scale=.45]{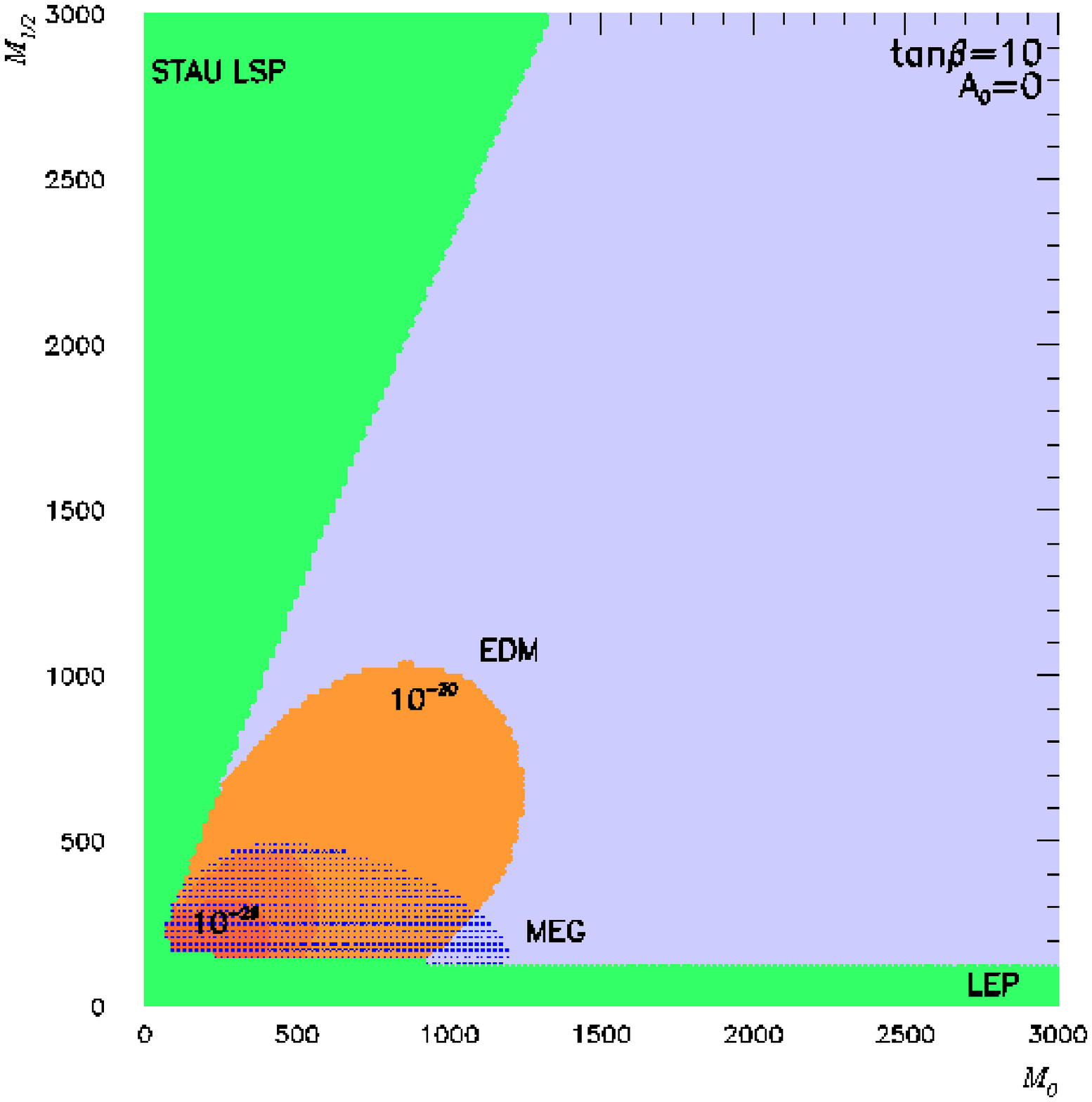} \includegraphics[scale=.45]{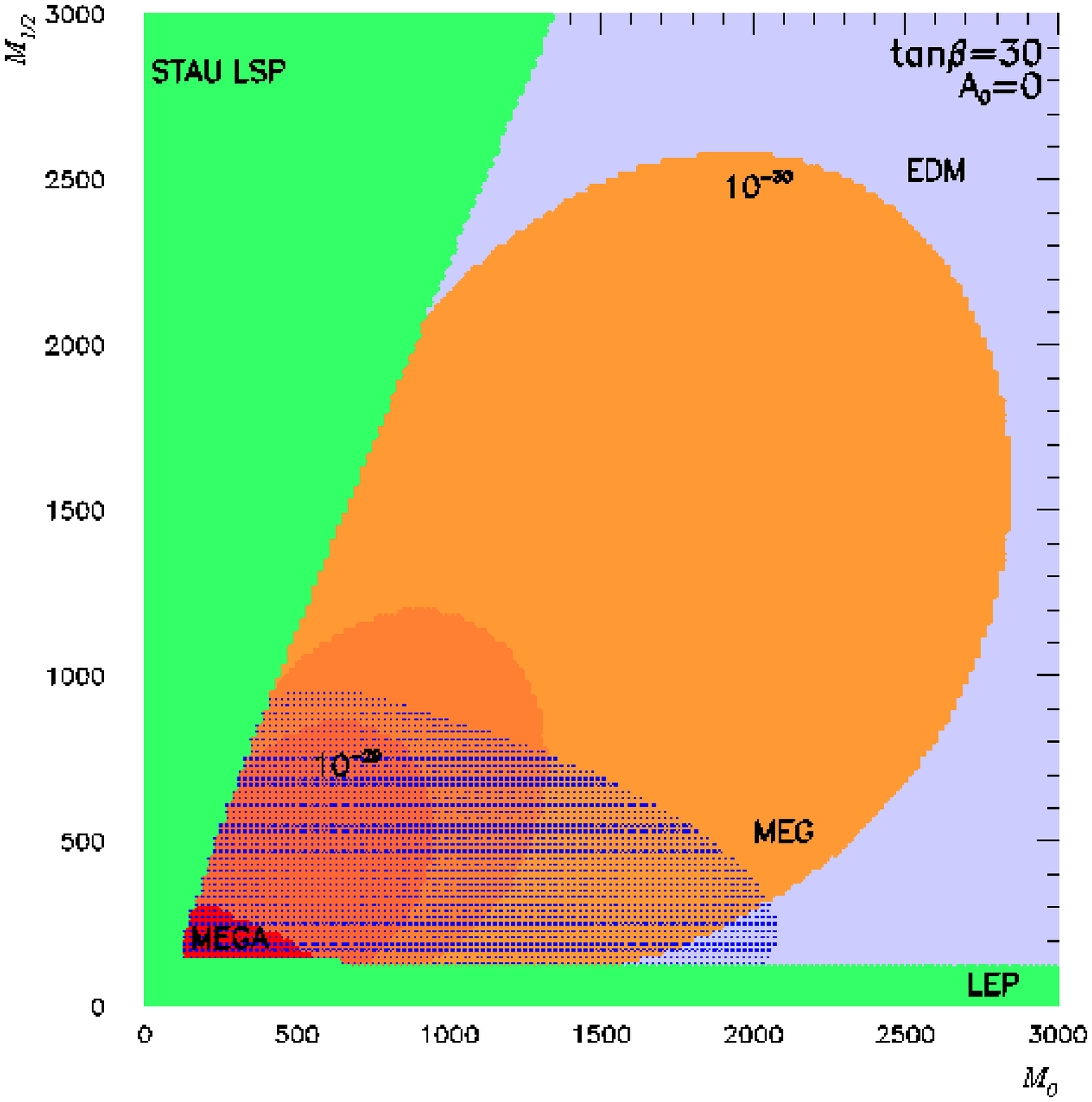}             
\caption{Values of $d_e$ in the $M_0$--$M_{1/2}$ plane for $\tan
  \beta =10,30$ and $A_0=0$. 
The hatched region corresponds to the reach of the
  future MEG experiment on $\mu\to e \gamma$ in the same model.}
\label{fig:EDM}
\end{figure}

As said above, in this $SU(3)$ flavour model CP violation is only broken
spontaneously by the flavon vevs below the Planck scale. In this way all terms
in the K\"ahler potential, giving rise to the soft masses and the $\mu$ term
by the Giudice-Masiero mechanism are real before the breaking of the flavour
symmetry. After breaking the flavour symmetry phases $O(1)$ will appear in the
Yukawa matrices and the off-diagonal elements of the soft mass matrices. In
this way $\mu$ is real before the breaking of the flavour symmetry. In fact,
even after the breaking of flavour and CP symmetries $\mu$ receives complex
corrections only at the two-loop level and therefore is still real to a 
very good approximation \cite{Ross:2004qn}. Similarly, 
diagonal elements in the trilinear terms are also real at leading order 
in the SCKM basis. In this way electric dipole moments (EDMs) are under
control and the SUSY CP problem is solved. Nevertheless off-diagonal phases in
the soft mass matrices contribute to the EDMs. For instance we have a
contribution to the electron EDM as
$d_e \propto m_\tau \mu \tan \beta \cdot {\rm Im }[ { 
\delta^{e_R}_{13}} \cdot {\delta^{e_L}_{31}}]$.
In figure \ref{fig:EDM} we show the expected contributions to the electron EDM
assuming that the phases in the off-diagonal elements are $O(1)$ and the
lepton Yukawas have CKM-like mixings \cite{Masiero:2002jn}. We can see
here that in this model, reaching a sensitivity of $10^{-29}$ e~cm in the
electron EDM will allow us to explore a significant region of the parameter
space even for intermediate values of $\tan \beta$ \cite{WIP}.      

\section{Conclusions}
The flavour problem in supersymmetric extensions of the SM is deeply related
to the origin of flavour in the Yukawa matrices. It is natural to think that the same 
mechanism generating the flavour structures in the Yukawa couplings is
responsible for the structure in the SUSY soft-breaking terms. In this way
finding a solution to the ``flavour problem'' in the SM can also provide a
solution to the SUSY flavour problem. In fact, the analysis of the 
new supersymmetric interactions can 
provide additional information on the physics of flavour which will be 
fundamental to improve our knowledge on flavour.
We have seen that measuring the flavour structures in the soft masses can help
us to ``measure'' the right-handed mixings in the Yukawa matrices. As an 
example, in an $SU(3)$ flavour model where the SUSY CP problem is also solved, 
we have shown the expected values for the electron EDM associated with
flavour non-diagonal SUSY phases.

%
%

\end{document}